\documentclass[onecolumn,secnumarabic,amssymb, nobibnotes, aps, prd,preprint]{revtex4-1}
\usepackage{color}

\usepackage{amsmath,amssymb}
\usepackage{bm}
\usepackage{graphicx}

\begin{document}

\title{Optically probing Schwinger angular momenta in a micromechanical resonator}%

\author{Motoki Asano$^{1,*}$}%
\author{Ryuichi Ohta$^1$}%
\author{Takuma Aihara$^2$}%
\author{Tai Tsuchizawa$^2$}%
\author{Hajime Okamoto$^1$}%
\author{Hiroshi Yamguchi$^1$}%

\affiliation{$^1$NTT Basic Research Laboratories, NTT Corporation, 3-1 Morinosato Wakamiya, Atsugi-shi, Kanagawa 243-0198, Japan}
\affiliation{$^2$NTT Device Technology Laboratories, NTT Corporation, 3-1 Morinosato Wakamiya, Atsugi-shi, Kanagawa 243-0198, Japan}

\date{\today}%

\begin{abstract}
We report an observation of phononic Schwinger angular momenta, which fully represent two-mode states in a micromechanical resonator. This observation is based on simultaneous optical detection of the mechanical response at the sum and difference frequency of the two mechanical modes. A post-selection process for the measured signals allows us to extract a component of phononic Schwinger angular momenta. It also enables us to conditionally prepare two-mode squeezed (correlated) states from a randomly excited (uncorrelated) state. The phononic Schwinger angular momenta could be extended to high-dimensional symmetry (e.g. SU(N) group) for studying multipartite correlations in non-equilibrium dynamics with macroscopic objects.
\end{abstract}
\maketitle

In the theory of quantum angular momentum generalized by Julian Schwinger \cite{schwinger}, the angular momentum constructed from two bosonic modes is classified into two groups, SU(2) and SU(1,1). Such Schwinger angular momenta fully represent two-mode states in the system in terms of energy, coherence, and correlations, and allow us to perform state tomography \cite{white,rehacek}, interferometry \cite{yurke}, and test of quantumness \cite{nha}. State tomography with angular momentum in the SU(2) group (i.e., Stokes parameters \cite{stokes}) has been widely demonstrated to identify coherence between various two-mode optical systems \cite{milione,bowen,lassen} in both classical and quantum regime. Interferometry with the angular momenta in the SU(1,1) group has a great potential to enable phase measurements at the Heisenberg limit \cite{li}. Such Schwinger angualr momenta have been well-engineered mainly in optical (photonic) systems but have never been explicitly observed in mechanical (phononic) systems.

A high-Q micromechanical resonator contains multiple non-degenerated mechanical modes and easily shows nonlinearlity induced by opto-mechanical or electro-mechanical interaction 
\cite{sankey,brawley,leijssen,okamoto,faust,mahboob,pontin}. The combination of such nonlinearity with the multiplicity in the mechanical modes opens the door to the observation of phononic Schwinger angular momentum in a mechanical system. In this Letter, we demonstrate optical probing of Schwinger angular momenta in a high-Q micromechanical resonator. The phononic Schwinger angular momenta in the SU(2) and SU(1,1) groups are revealed by probing nonlinear signals via Doppler interferometry and performing a post-selection process. They enable us to conditionally extract two-mode squeezed (correlated) states from a randomly excited (uncorrelated) state because these angular momenta are the function of cross-correlation.

The phononic Schwinger angular momenta in the SU(2) and SU(1,1) groups are represented by vectors ${\bm S}=(S_x, S_y, S_z)$ and ${\bm K}=(K_x, K_y, K_z)$ in a sphere and a hyperboloid, respectively [Fig. 1(a) and (b)]. The vertical axis of the sphere is given by $S_z=(x_2^2+y_2^2-x_1^2-y_1^2)/4$ and that of the hyperboloid is given by $K_z=(x_2^2+y_2^2+x_1^2+y_1^2)/4$, where $x_i$ and $y_i$ are the linear quadrature of $i$th mechanical modes. The $z$ component of the Schwinger angular momentum corresponds to the energy difference between the two mechanical modes for the SU(2) group and the energy sum for the SU(1,1) group. On the other hand, the $x$ and $y$ components, $S_x$, $S_y$ and $K_x$, $K_y$, include information about the cross-correlation of the two mechanical modes, where $S_x=(x_2y_1-y_2x_1)/2$, $S_y=-(x_2x_1+y_1y_2)/2$, $K_x=(x_2y_1+x_1y_2)/2$, and $K_y=(x_2x_1-y_2y_1)/2$.  Each component of angular momenta satisfies a Lie algebra, e.g., $\{S_i,S_j\}={S}_k$ only if $i$, $j$, $k$ are cyclic with $x$, $y$, $z$ with the Poisson bracket $\{\cdot\}$ (see supplemental information). These angular momenta represent rotational trajectories in their joint phase space. The rotational trajectory in the SU(2) group is specified with a real angle and that in the SU(1,1) group is specified with an imaginary angle (i.e., hyperbolic trajectory). For instance, the phononic Schwinger angular momentum along the $x$-direction, i.e., ${\bm S}=(S_x,0,0)$ (${\bf K}=(K_x,0,0)$), represents the rotational trajectory with a real angle $\phi$ (an imaginary angle $i\phi$) in a joint phase space spanned by $x_1$ and $x_2$.
\begin{figure}[htb]
\includegraphics[width=70mm]{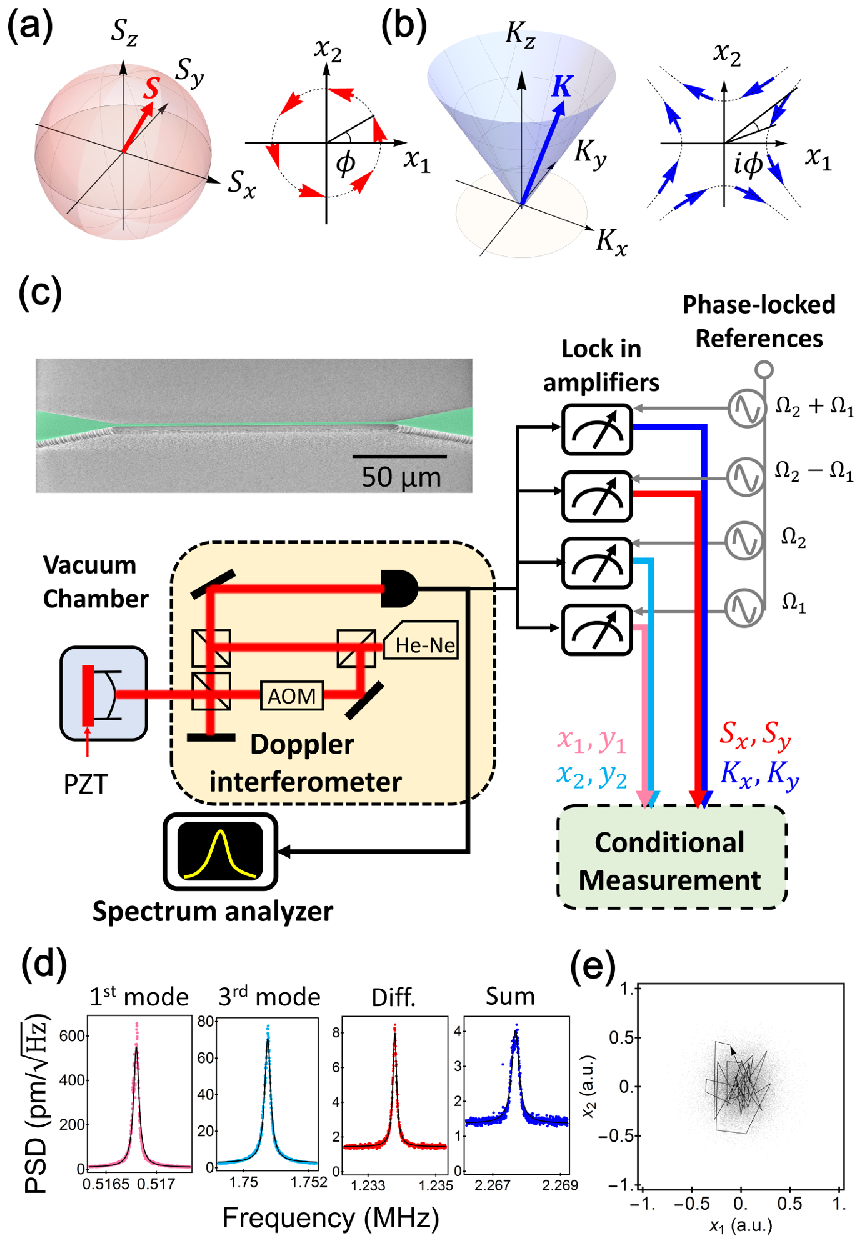}
\caption{\label{Fig1}(a) and (b) Illustrations of spherical and pseudo-spherical (hyperboloid) space with vectors ${\bm S}$ and ${\bm K}$ of angular momentum in SU(2) and SU(1,1) groups. The typical rotational trajectories in the joint phase space with ${\bm S}=(S_x, 0, 0)$ and ${\bm K}=(K_x, 0, 0)$ are described with a real and imaginary angle, $\phi$ and $i\phi$, in their insets. (c) Schematic of the experimental setup at room temperature. A doubly-clamped mechanical beam was fabricated from a high tensile (∼0.2 GPa) silicon-nitride film deposited by thermal chemical vapor deposition (see SEM image). A probe light from He-Ne laser with a wavelength of 633 nm was modulated by an acousto-optic modulator (80 MHz) and detected by an optical heterodyne setup. The detected signals were divided by two: one was connected to a spectrum analyzer to directly measure the power spectral density, the other one was connected to four lock-in amplifiers. Both the linear quadrature and transverse components of phononic Schwinger angular momenta in the SU(2) and SU(1,1) groups were detected by the lock-in amplifiers with phase-synchronized references. Each quadrature was used to perform a conditional measurement. (d) Power spectral density (PSD) of the first-order, third-order, difference frequency, and sum frequency signals measured in the spectrum analyzer. The vertical unit was calibrated to the displacement ($\mathrm{pm/\sqrt{Hz}}$) using the thermal motion of the first mechanical mode without random excitation. (e) Random trajectory (black vector) and distribution (black points) in the joint phase space spanned by $x_1$ and $x_2$.}
\end{figure}

Non-degenerated modes in a micromechanical resonator crucially contain randomness at finite temperature because they are independently actuated by random forces. They lead a random trajectory in their joint phase space, i.e., the phononic Schwinger angular momenta also show a random distribution. The average value of their transverse components becomes zero because of their isotropic distribution. In other words, we can achieve the two-mode mechanical states with the non-zero average of phononic Schwinger angular momenta by directly probing these components and conditionally extracting part of the distribution. To demonstrate these operations, we perform the following two steps.

The first step is to directly probe the nonlinear optical signals generated via higher-order modulation in a Doppler interferometer. These nonlinear optical signals at the sum and difference frequency of two mechanical modes correspond to the response from the transverse components of the phononic Schwinger angular momenta in the SU(2) and SU(1,1) groups, respectively (see supplemental information). A Doppler interferometer was used to simultaneously probe the first-order vibrational mode ($\Omega_1=2\pi\times 0.52$ MHz) and third-order vibrational mode ($\Omega_2=2\pi\times 1.75$ MHz) in a high-tensile silicon-nitride mechanical resonator (150-$\mathrm{\mu}$m long, 5-$\mathrm{\mu}$m wide, and 525-nm thick) [Fig. 1(c)].  The high quality factors ($\sim30,000$) in our resonator enhance the measurement sensitivity of the sum and difference frequency signals because the power spectral density of them is proportional to $\sqrt{n_1n_2Q}$, where $n_i$ is the number of phonons in the $i$th mode and $Q$ is the quality factor (see supplemental information). To overcome the measurement noise ($\sim$ 1.4 $\mathrm{pm/\sqrt{Hz}}$) in our setup, the mechanical modes were additionally excited to increase $n_1$ and $n_2$ via a PZT sheet with artificial white noise. Around the effective temperature $T_\mathrm{eff}>10^6$ K, we observed the noise spectra of sum and difference frequency signals with a spectrum analyzer (the vertical unit is calibrated with thermal motion at room temperature without excitation \cite{karabalin})[Fig. 1(d)]. A random trajectory in the joint phase space was obtained  by monitoring the linear quadrature with two lock-in amplifiers with $\Omega_1$ and $\Omega_2$ frequency references [Fig. 1(e)].

The second step involves post-selecting the non-zero average in the transverse components of phononic Schwinger angular momenta. This conditional measurement was performed in quadrature of nonlinear signals from two additional lock-in amplifiers with $\Omega_2-\Omega_1$ and $\Omega_2+\Omega_1$ frequency references. This quadrature contains the response from the two orthogonal transverse components of phononic Schwinger angular momenta (e.g., $S_x$ and $S_y$). The $10^5$ temporal data sets for all signals from the four lock-in amplifiers were recorded within a total of 400 msec. To extract the statistical ensemble which shows the non-zero average of the transverse components, we imposed a condition $O_j/\mathrm{max}[O_j]\geq \eta$ for a quadrature (i.e., $O_j$=$S_x$, $S_y$, $K_x$ and $K_y$), where $\eta$ is a trigger level [Fig. 2(a)]. Although increasing $\eta$ allows us to extract a large average value of $O_j$, the number of events which satisfy the condition decreases. We evaluate success probability $p_s$ defined by the number of events satisfying the condition divided by the total number of events [Fig. 2(b)].  When $\eta=0$, only the positive value of the angular momentum is extracted from the random motion, and the average value is trivially non-zero, where $p_s=0.5$. Increasing $\eta$ decreases the success  probability because a large amount of the angular momentum only slightly appears in the random motion. The trajectory in the conditional measurement was determined by extracting  linear quadratures $x_i$ and $y_i$ only when $O_j$ satisfies the condition. To display the rotational trajectories with respect to the phononic Schwinger angular momenta, the reference frequencies for the lock-in amplifiers were set with a finite detuning so that the detuning $\delta_i$ of the $i$th mode satisfies $\delta_1\approx \delta_2$ (see supplemental information). The conditional measurement with the finite detuning results in discontinuous rotational motion with a real and imaginary angle [conceptually depicted in Fig.1(a) and (b)] with $S_x$ and $K_x$, respectively [Fig. 2(c) and 2(d)], whereas a random trajectory was observed without any conditioning [Fig. 1(e)]. This verifies that our conditional measurement allows us to extract the non-zero average of phononic Schwinger angular momenta. For the other components, $S_y$ and $K_y$, we achieved similar results in the phase space spanned by $x_1$ and $y_2$ (see supplemental information). 
\begin{figure}
\includegraphics{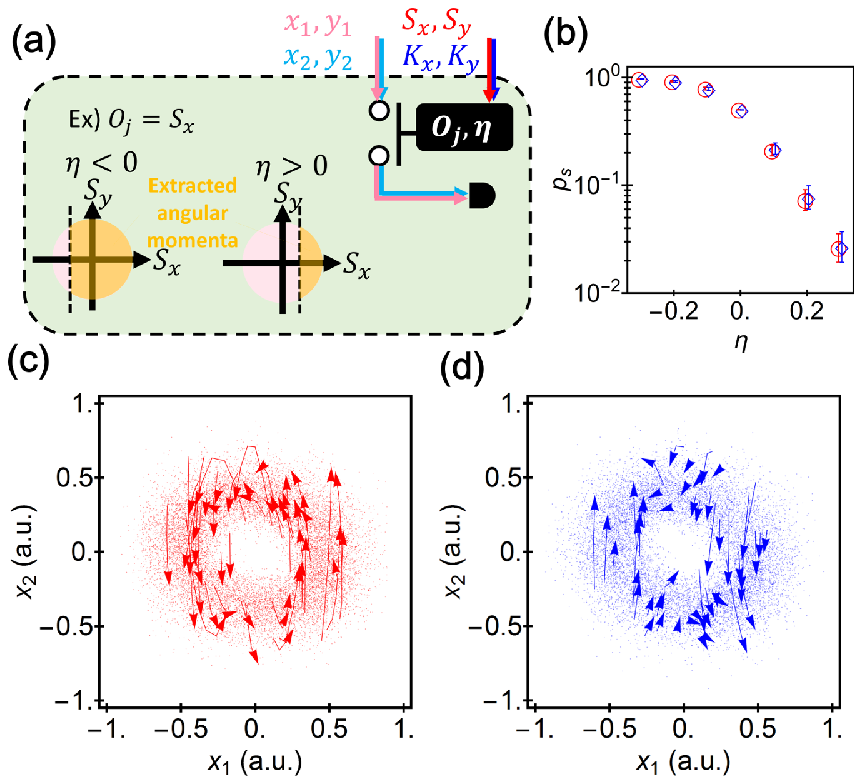}
\caption{(a) Conceptual illustration of conditional measurement with the input of the linear quadrature of mechanical modes and transverse components of phononic Schwinger angular momenta. Two typical conditions with positive and negative $\eta$ are depicted. (b) Success probability with respect to $\eta$ for $S_x$ (the red circles) and $K_x$ (the blue diamonds). The error-bar shows the standard deviation for ten trials with the same condition. (c) and (d) Extracted trajectory (vectors) and distributions (points) with $\eta=0.3$ for $S_x$ and $K_x$, respectively. The discontinuous trajectories are described while the Schwinger angular momenta satisfy the condition.}
\end{figure}

\begin{figure}
\includegraphics{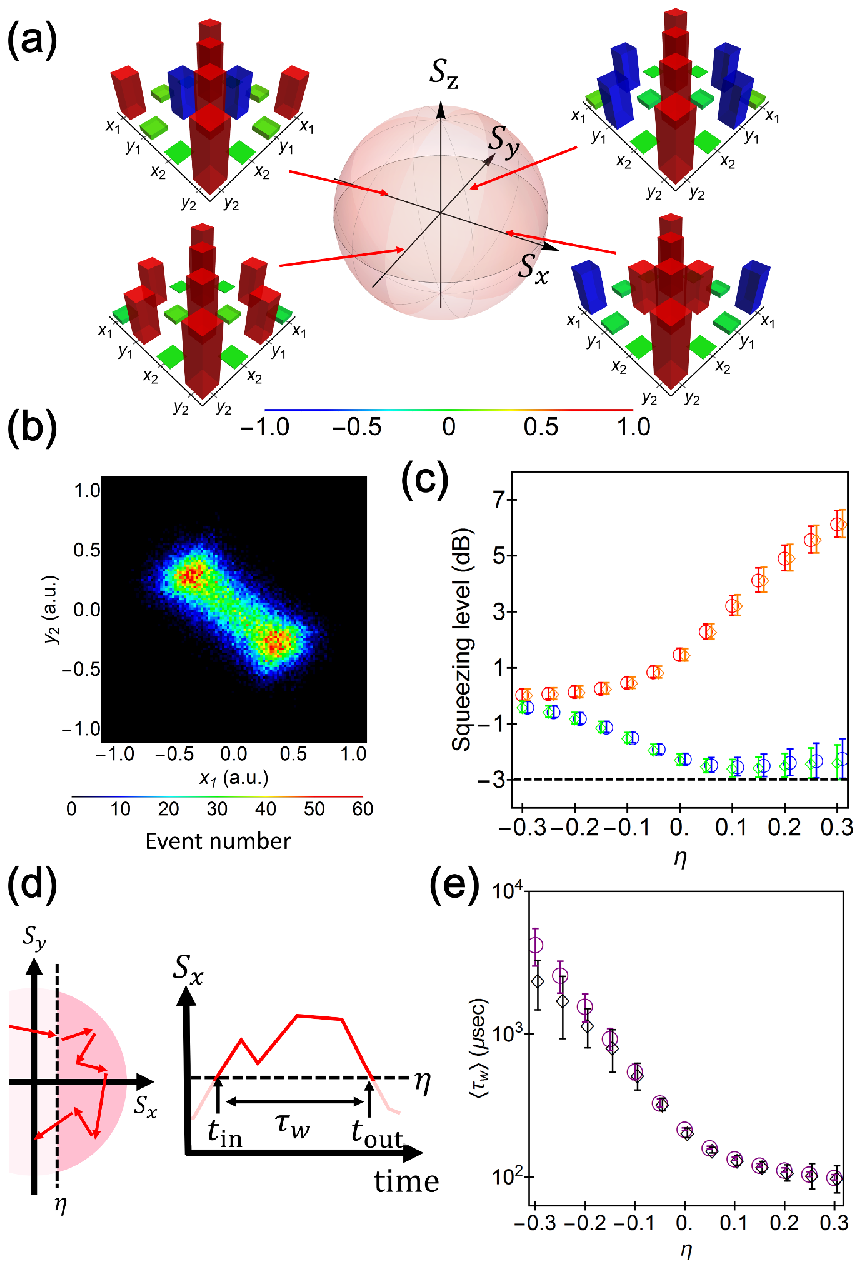}
\caption{(a) Correlation coefficient matrices with conditional measurements when $O_j=\{S_x, S_y\}$ and $\left|\eta\right|=0.2$ in Schwinger angular momentum in SU(2) group. The absolute value and sign of the correlation coefficient are described by the color maps. (b) Density histogram of the extracted two-mode squeezed state in the joint phase space spanned by $x_1$ and $y_2$ with $\eta=0.2$ for $S_x$. (c) Squeezing level with different $\eta$. The red, orange, green and blue plots show the noise reduction ratio along diagonal axes $x_1-y_2$, $y_1+x_2$, $y_1-x_2$, and $y_1+x_2$, respectively. The black dashed line corresponds to the squeezing level of -3 dB. (d) Schematic image of the survival-time duration. The red vectors describe the random motion projected to the space spanned by $S_x$ and $S_y$. (e) Average of the survival time duration with respect to $\eta$. The numerical results show the good agreement with the experimental time duration. The purple and black plots corresponds to the experimental and numerical results. All error-bars were calculated as the standard deviation for ten trials.}
\end{figure}

The transverse components of phononic Schwinger angular momenta also represent a correlation between the two linear quadratures. For instance, the average of $S_x$ and $K_x$ contains the correlation function of $x_1$ and $y_2$ (also $x_2$ and $y_1$). This implies that the conditional measurement  allows us to extract a correlated two-mode state in the randomly actuated mechanical resonator. Here, we evaluate the two-mode correlation in the extracted ensemble with a correlation coefficient matrix. Each element is defined by the correlation factor $C_{ij}=\mathrm{Cov}[u_i,u_j]/(\sigma(u_i)\sigma(u_j))$ ($u_i=\{x_i,y_i\}$), where $\mathrm{Cov}[x,y]$ denotes the covariance between $x$ and $y$, and $\sigma_i (x)$ denotes the standard deviation of $x$. For instance, only self-correlation represented in the diagonal elements appears in the matrix without any conditioning. On the other hand, cross correlation represented in the off-diagonal elements appears with the condition in the phononic Schwinger angular momentum in the SU(2) group [Fig. 3(a)]. Note that the negative average was extracted with the condition $O_j/\mathrm{max}⁡[O_j ]\leq -\eta$ with a positive $\eta$. There totally exist eight types of correlation including the combination of the linear quadrature and the sign of the correlation (four residual correlations were obtained in angular momentum in the SU(1,1) group [see supplemental information]). Moreover, such correlated states show the reduction and amplification of the noise deviation along 45-degree and -45-degree axes in the phase space [Fig. 3(b)]. Increasing $\eta$ decreases the squeezing level defined by the ratio of the standard deviation between the extracted ensemble and the ensemble without any conditioning [Fig. 3(c)]. The lower limit of the squeezing level reached -3 dB as the two-mode squeezing via parametric nonlinearity does \cite{mahboob,pontin}. This means that the conditional measurement of a transverse component suppresses half of the two-mode noise in the phase space. Because the random dynamics of the mechanical modes is determined by a finite time constant, the two-mode squeezed state can survive in the conditional window with a finite time duration $\tau_w\equiv t_\mathrm{out}-t_\mathrm{in}$, where $t_\mathrm{in}$ ($t_\mathrm{out}$) is the time when the Schwinger angular momenta enters (leaves) the conditioned window [Fig. 3(d)]. The average of the time duration $\langle \tau_w\rangle$ tells us how fast we should perform the operation by leveraging the squeezed states. Apparently, it exponentially decreases with increasing $\eta$ [Fig. 3(e)] \cite{note}. 

\begin{figure}
\includegraphics{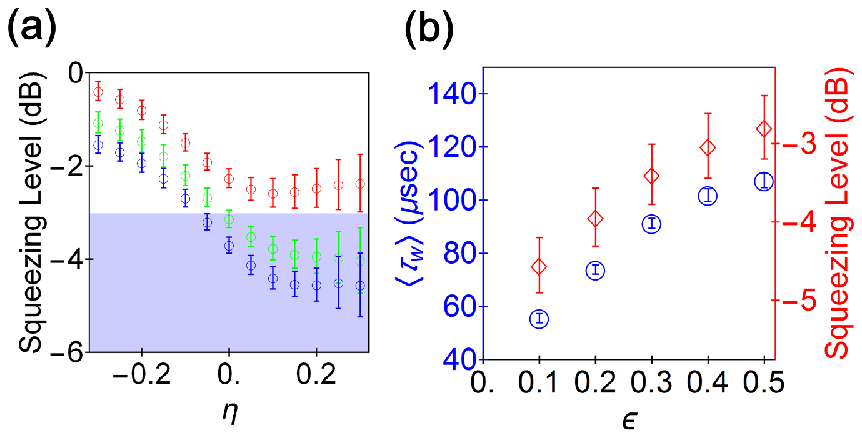}
\caption{(a) Squeezing level with respect to $\eta$ and three different $\epsilon$ with $O_j=S_x$ and $\tilde{O_j}=S_y$. The red plots show the level with the single conditional measurement, and the blue and green plots show the case with $\epsilon=0.2$ and $\epsilon=0.1$. The blue-shaded area corresponds to the level over -3 B that is never obtained in the single conditioning.  (b) Average of the survival-time duration (blue plots) and squeezing level (red plots) with $\eta=0.2$ and different $\epsilon$. All error-bars were calculated as the standard deviation for ten trials.}
\end{figure}
Further suppressing the deviation of the Schwinger angular momentum can improve the level of two-mode squeezing. We imposed the second condition, $-\epsilon\leq \tilde{O}_j /\mathrm{max⁡}[\tilde{O}_j]\leq \epsilon$, where $\tilde{O}_j$ is the residual component of the SU(2) [or SU(1,1)] angular momentum, and $\epsilon$ is the positive parameter that determines the window for decreasing the appearance of the residual component. Here, we show the case where $O_j=S_x$ and $\tilde{O_j}=S_y$. The squeezing level of the extracted states is improved to over -3 dB with decreasing $\epsilon$ [Fig. 4(a)]. There exists a trade-off between the squeezing level and the average of the survival-time duration $\langle \tau_w\rangle$ with decreasing $\epsilon$ as well as $\eta$ [Fig. 4(b)]. Note that the squeezing level is saturated as it approaches -6 dB because the deviation along $S_z$ still exists, and it cannot be accessed in our scheme. Such a double-conditioning scheme can be readily extended to an arbitrary set of the two conditions for both the Schwinger angular momenta in the SU(2) and SU(1,1) groups. Although the other conditions enable us to extract the various ensembles, the squeezing level does not exceed -3 dB (see supplemental information). 

In conclusion, we observed phononic Schwinger angular momenta in SU(2) and SU(1,1) groups between two mechanical modes by detecting nonlinear optical signals and performing a conditional measurement. This conditional measurement also enables us to extract two-mode correlated (squeezed) states. The concept of Schwinger angular momenta is available for representing not only two-body system but also many-body ones with a higher-order symmetry (e.g., that in the SU(N) group \cite{mathur}).  Such high-dimensional angular momenta can be probed with state-of-the-art cavity nonlinear optomechanical systems \cite{sankey,brawley,leijssen} integrated with multimode mechanical structures \cite{mahboob2,hvidtfelt}.  They open the way to the study of multipartite correlations in non-equilibrium many-body systems affected by various kind of heat baths \cite{groblacher, klaers}. Furthermore, as in the scheme of continuous quantum measurement in a mechanical oscillator \cite{jacobs, vanner}, probing the phononic Schwinger angular momenta in a few-phonon regime could be an interesting challenge for the study of non-equilibrium thermodynamics with macroscopic quantum objects.

We thank Kensaku Chida, Yuichiro Matsuzaki, Tatsuro Hiraki, and Samer Houri for fruitful discussions. This work was partly supported by a MEXT Grant-in-Aid for Scientific Research on Innovative Areas (Grants No. JP15H05869).

\clearpage

\pagebreak
\widetext
\begin{center}
\textbf{\large Supplemental Information: Optically probing Schwinger angular momenta in a micromechanical resonator}
\end{center}

\setcounter{equation}{0}
\setcounter{figure}{0}
\setcounter{table}{0}
\setcounter{page}{1}
\makeatletter
\renewcommand{\theequation}{S\arabic{equation}}
\renewcommand{\thefigure}{S\arabic{figure}}
\renewcommand{\bibnumfmt}[1]{[S#1]}

\section*{Schwinger angular momenta between two bosonic modes}
The components of Schwinger angular momenta $S_i $ in SU(2) and $K_i$ in SU(1,1) ($i=x,y,z$) are defined as quadratic forms of linear phase quadratures $x_j$ and $y_j$ ($j=1,2$) as follows:
\begin{align}
S_x=&(x_2y_1-y_2x_1)/2,\\
S_y=&-(x_2x_1+y_1y_2)/2,\\
S_z=&(x_2^2+y_2^2-x_1^2-y_1^2)/4,\\
K_x=&(x_2y_1+x_1y_2)/2,\\
K_y=&(x_2x_1-y_2y_1)/2,\\
K_z=&(x_2^2+y_2^2+x_1^2+y_1^2)/4,
\end{align}
Because $x_i$ and $y_i$ are canonical conjugate with each other, the algebra among the angular momenta can be examined by taking account of the Poisson bracket $\{A,B\}\equiv \sum_{i=1}^2 \frac{\partial A}{\partial x_i}\frac{\partial B}{\partial y_i}-\frac{\partial A}{\partial y_i}\frac{\partial B}{\partial x_i}$. For $S_x$ and $S_y$, we obtain the following cyclic algebra
\begin{align}
\{S_x,S_y\}=S_z \hspace{5pt},\{S_y,S_z\}=S_x \hspace{5pt},\{S_z,S_x\}=S_y,\\
\{K_x,K_y\}=-K_z \hspace{5pt},\{K_y,K_z\}=K_x \hspace{5pt},\{K_z,K_x\}=K_y.
\end{align}
The $z$ components of SU(2) [SU(1,1)] angular momentum commute with the transverse components of SU(1,1) [SU(2)] angular momentum as
\begin{align}
\{S_x, K_z\}=\{S_y, K_z\}=\{K_x, S_z\}=\{K_y, S_z\}=0.
\end{align}
Moreover, the combination of $S_i$ and $K_j$ ($i,j=x,y$) is no longer closed in the algebra of angular momentum, such that
\begin{align}
\{S_x, K_x\}=&(x_1y_1-x_2y_2)/2\\
\{S_x, K_y\}=&-(x_1^2-y_1^2-x_2^2+y_2^2)/4\\
\{S_y, K_x\}=&-(x_1^2-y_1^2+x_2^2-y_2^2)/4.\\
\{S_y, K_y\}=&-(x_1y_1+x_2y_2)/2\\
\end{align}
The observables in the right-hand-side are achieved in the second harmonics of mechanical modes from the measurement nonlinearity (experimentally observed in \cite{brawley}). All observables including the second harmonics and the Schwinger angular momenta constructs $sp(4)$-algebra, as discussed in \cite{vourdas}. Note that the quantization of these angular momenta immediately derives the commutation relationship including the Planck constant with the same algebra.

\section*{Phononic Schwinger angular momenta in Doppler effect}
The Doppler effect causes the frequency modulation of light with respect to the velocity of vibrational objects. Here, we consider that the light simultaneously probes two mechanical vibrations with different angular frequencies, $\Omega_1$ and $\Omega_2$, at finite temperature. In our heterodyne setup, the probe light $a(t)$ is given by
\begin{align}
a(t)=\sqrt{P} e^{i\Omega_\mathrm{AOM}t}\exp\left[i(\beta_1v_1(t)+\beta_2v_2(t))\right]+c.c.
\end{align}
where $P$ is the probe power, $\Omega_\mathrm{AOM}$ is the angular frequency shifted by the acousto-optic modulator, and $\beta_i$ and $v_i$ are the modulation index and  the vibrational velocity of $i$th mechanical mode. The velocity driven by random force is represented by $v_i(t) =x_i \cos\Omega_i t+y_i \sin\Omega_i t$, where $\Omega_i$ is the angular frequency, $x_i(t)$ and $y_i(t)$ are orthogonal linear quadratures. Using Jacobi-Anger expansion, we obtain
\begin{align}
a(t)=&\sqrt{P} e^{i\Omega_\mathrm{AOM}t}\sum_k \sum_l\sum_m \sum_n  i^{(k+l)} J_k\left(\beta_1 x_1\right)J_l\left(\beta_2 x_2\right)\nonumber\\
&J_m\left(\beta_1 y_1\right)J_n\left(\beta_2 y_2\right) e^{i((k+m)\Omega_1+(l+n)\Omega_2)t}+c.c.,
\end{align}
where $J_m(x)$ is the $m$th order Bessel function of the first kind. This can be expanded in terms of optical sideband frequencies (including the difference frequency  $\Omega_D\equiv \Omega_2-\Omega_1$ and sum frequency $\Omega_S\equiv \Omega_2+\Omega_1$) as follows:
\begin{align}
a(t)=&\sqrt{P}\Biggl[a_0(t)+a_{\Omega_\mathrm{AOM}\pm\Omega_1}(t)+a_{\Omega_\mathrm{AOM}\pm\Omega_2}(t)\nonumber\\
&+a_{\Omega_\mathrm{AOM}\pm\Omega_D}(t)+a_{\Omega_\mathrm{AOM}\pm\Omega_S}(t)+\dots\Biggr]
\end{align}
where each term is given by
\begin{align}
a_{\Omega_\mathrm{AOM}\pm\Omega_i}(t)=&\beta_i x_i(t)\sin\left[(\Omega_\mathrm{AOM}\pm\Omega_i)t\right]\nonumber\\
&\pm \beta_i y_i(t)\cos\left[(\Omega_\mathrm{AOM}\pm\Omega_i)t\right]\\
a_{\Omega_\mathrm{AOM}\pm\Omega_D}(t)=&\pm\beta_1\beta_2 S_x(t)\sin\left[(\Omega_\mathrm{AOM}\pm\Omega_D)t\right]\nonumber\\
&+\beta_1\beta_2S_y(t)\cos\left[(\Omega_\mathrm{AOM}\pm\Omega_D)t\right]\\
a_{\Omega_\mathrm{AOM}\pm\Omega_S}(t)=&\mp\beta_1\beta_2 K_x(t)\sin\left[(\Omega_\mathrm{AOM}\pm\Omega_S)t\right]\nonumber\\
&-\beta_1\beta_2K_y(t)\cos\left[(\Omega_\mathrm{AOM}\pm\Omega_S)t\right].
\end{align}
Here, we assume that $\beta_i x_i \ll 1$ and $\beta_i y_i\ll 1$ for approximating the Bessel function as $J_q(z)\approx (z/2)^q/q!$. Obviously, the difference (sum) frequency signal contains the transverse components of phononic Schwinger angular momentum in the SU(2) [SU(1,1)] group.

\section*{Signal level for phononic Schwinger angular momenta}
To discuss the signal level of phononic Schwinger angular momenta, we introduce the power spectral density, which is defined by the Fourier transform of the statistical average, $\langle z(t)z(0)\rangle$, of thermal noise at the mechanical frequency:
\begin{align}
S_{v_i}(\Omega_i)=\sqrt{\frac{2k_B T}{m_i\Gamma_i}}=2\sqrt{n_i Q_i \Omega_i}x_i^\mathrm{zp} \hspace{10pt}(i=1,2)
\end{align}
with a unit of (m/s)/$\sqrt{\mathrm{Hz}}$ where $k_B$ is the Boltzmann factor, $T$ is the temperature, and $m_i$, $\Gamma_i$, $n_i$, $Q_i$, and $x_i^\mathrm{zp}$ are the effective mass, linewidth, number of thermal phonons, quality factor, and zero-point fluctuation of mechanical displacement of the $i$th mechanical mode, respectively. In the same manner, the sensitivity of $O_j=\{S_x, S_y, K_x, K_y\}$ is represented by
\begin{align}
S_{O_j}(\Omega_i)=&\sqrt{\frac{2(k_BT)^2}{m_1m_2\Omega_1^2\Omega_2^2(\Gamma_1+\Gamma_2)}}\nonumber\\
&=2\sqrt{2}\sqrt{\frac{n_1n_2Q_1Q_2\Omega_1^2\Omega_2^2}{\Omega_1Q_2+\Omega_2Q_1}}x_1^\mathrm{zp}x_2^\mathrm{zp} \hspace{10pt}(i=D,S)
\end{align}
with a unit of $\mathrm{(m/s)}^2$/$\sqrt{\mathrm{Hz}}$. Here, we assume that $\langle x_i(t)x_i(0)y_i(t)y_i(0)\rangle=\langle x_i(t)x_i(0)\rangle\langle y_i(t)y_i(0)\rangle$. Because the modulation index in the Doppler effect is represented by $\beta_i=2\pi /(\lambda_L\Omega_i)$, where $\lambda_L$ is the wavelength of the probe light, the power spectral densities of measured photocurrent for linear and nonlinear terms are given by
\begin{align}
I_\mathrm{L}(\Omega_i)=& 4\pi \xi_D x_i^\mathrm{zp} \sqrt{\frac{PP_L n_i Q_i}{\Omega_i  \lambda_L^2}}\hspace{10pt}(i=1,2)\\
I_\mathrm{NL}(\Omega_i)=&8\sqrt{2}\pi^2\xi_Dx_1^\mathrm{zp}x_2^\mathrm{zp} \sqrt{\frac{PP_Ln_1n_2Q_1Q_2}{(\Omega_1Q_2+\Omega_2Q_1)\lambda_L^4}}\nonumber\\
&\hspace{100pt}(i=D,S)
\end{align}
where $P_L$ is the power of the local oscillator, $\xi_D$ is the factor of opto-electric conversion in the photodetector. To discuss the signal-to-noise ratio for nonlinear signals, the nonlinear power spectral density $\tilde{S}_{O_j}(\Omega_i)$ converted as the unit of $\mathrm{(m/s)/\sqrt{Hz}}$ is defined by
\begin{align}
\tilde{S}_{O_j}(\Omega_i)\equiv & \frac{I_\mathrm{NL}(\Omega_i)}{I_\mathrm{L}(\Omega_1)} S_1(\Omega_1).
\end{align}
\begin{table}
\caption{Parameters for power spectral density}
\begin{ruledtabular}
\begin{tabular}{lll}
Symbol& Value & Unit \\ \hline\hline 
$\Omega_1$ & $2\pi\times 0.52$ & MHz \\
$\Omega_2$ & $2\pi\times 1.74$ & MHz \\
$x_1^\mathrm{ZP}$ & 5.6 & fm\\
$x_2^\mathrm{ZP}$ & 2.9 & fm\\
$Q_1, Q_2$& $3.0 \times 10^4$\\
$\lambda_L$& 632.8 & nm
\end{tabular}
\end{ruledtabular}
\end{table}
In our experiment, the noise floor level in the Doppler interferometer was estimated to 1.4 $\mathrm{pm}/\mathrm{\sqrt{Hz}}$. This value was calibrated by a thermal noise spectrum at room temperature in the first mechanical mode. The nonlinear signals were obtained when the effective temperature $T_\mathrm{eff}\equiv n_1\hbar\Omega_1/(k_B T)\sim 10^6$ with the artificial noise amplitude $V_\mathrm{noise}\sim 2$ V. The experimental power spectral density shows good agreement with the theoretical value [Fig. S1(a)] calculated with the parameters in Table I. Mechanical resonators with a high Q factor and small effective mass enable us to observe the Schwinger angular momenta at room temperature without any artificial noise [Fig. S1(b)]. For instance, mechanical resonators with two-dimensional materials \cite{bunch, will} and nanowire mechanical resonators \cite{abhilash, he2} have extremely small effective mass, and silicon-nitride resonators with a phononic shield \cite{tsaturyan, ghadimi} show $Q>10^6$. They will be good candidates for observing and engineering phononic Schwinger angular momenta. 
\begin{figure}
\includegraphics{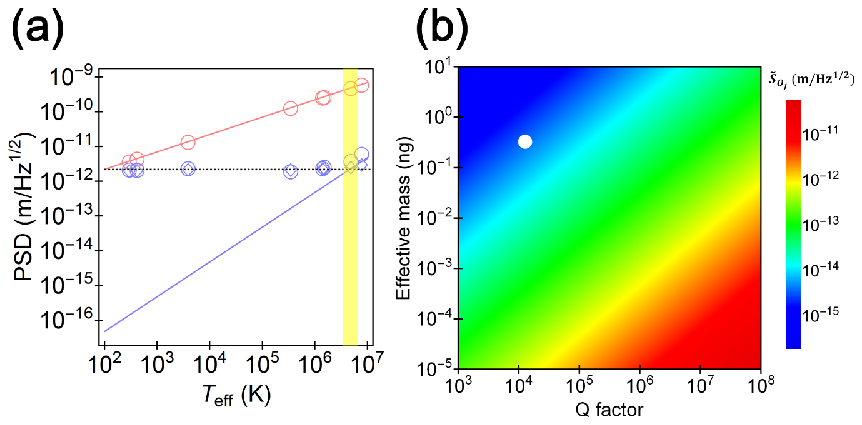}
\caption{(a) Power spectral density of the first mechanical modes (red circle) and the sum and difference frequency signals (blue circle: sum frequency signal, and blue square: difference frequency signal). The red and blue lines show the theoretically calculated values with the parameters in Table I. The effective temperature $T_\mathrm{eff}$ is calibrated with the first mechanical modes. The yellow-shaded area corresponds to $V_\mathrm{noise}=2$ V, at which we performed all experiments shown in the main text. The black dashed line shows the noise floor level in the Doppler interferometer. (b) Density color plots of the converted nonlinear power spectral density as a function of mechanical Q factor and effective mass at room temperature (300 K). The white point is the parameter in our experiment.}
\end{figure}

\section*{Rotational trajectory with a finite detuning}
The dynamics of two-mode mechanical system in a rotational reference frame is represented by the Hamiltonian $\mathcal{H}=\sum_i \delta_i(x^2_i+y_i^2)/2$, where $\delta_i$ is a detuning between the frequency of the reference frame and resonance of the $i$th mode. By taking into account the random mechanical force, the linear quadratures follow the Langevin equations
\begin{align}
\dot{x}_i=&-\frac{\gamma_i}{2}x_i+\frac{\delta_i}{2}y_i+\sqrt{\gamma_i}x_{\mathrm{th},i}\\
\dot{y}_i=&-\frac{\gamma_i}{2}y_i-\frac{\delta_i}{2}x_i+\sqrt{\gamma_i}y_{\mathrm{th},i}\\
\end{align}
where $\gamma_i$ is the damping factor and $\xi_{\mathrm{th},i}$ ($\xi=x,y$) is the random fluctuation with $\langle \xi_{\mathrm{th},i}(t)\xi_{\mathrm{th},i}(t') \rangle=4k_BT_\mathrm{eff}\delta(t-t')$. 

Rotational trajectory in the joint phase space is specified by an angular momentum defined by the vector product of coordinates and their velocity fields (momentum). To discuss the detuning dependence of the rotational trajectory, here we define statistical averages of {\it rotational angular momenta} $\mathcal{S}\equiv\langle\dot{x}_2x_1-\dot{x}_1x_2\rangle_\eta$ and $\mathcal{K}\equiv\langle \dot{x}_2x_1+\dot{x}_1x_2\rangle_\eta$, where $\langle\cdot\rangle_\eta$ denotes the ensemble average with the conditional measurement with $\eta$. $\mathcal{S}$ quantifies the rotational trajectory with a real angle, and $\mathcal{K}$ does so with a imaginary angle.
By considering the conditional measurement with $S_x$ and $K_x$ (i.e. $\langle S_y\rangle_\eta=\langle K_y\rangle_\eta=0$) and substituting the Langevin equations into the definition of the rotational angular momenta, we obtain relationships
\begin{align}
\mathcal{S}=-\frac{\delta_2+\delta_1}{2} \langle S_x\rangle_\eta +\frac{\delta_2-\delta_1}{2} \langle K_x\rangle_\eta,\label{Srot}\\
\mathcal{K}=\frac{\delta_2+\delta_1}{2} \langle K_x\rangle_\eta -\frac{\delta_2-\delta_1}{2} \langle S_x\rangle.\label{Krot}
\end{align}
A finite detuning crucially determines the rotational trajectory with respect to the conditional average of Schwinger angular momenta. In the case where $\delta=\delta_1=\delta_2$, the rotational trajectory with a real (imaginary) angle is determined by the phononic Schwinger angular momentum in the SU(2) [SU(1,1)] group. This corresponds to the fact that the phonoinc Schwinger angular momentum in the SU(2) [SU(1,1)] group is a generator of rotation with a real (imaginary) angle. $\mathcal{S}$ and $\mathcal{K}$ obtained in numerical simulation of the Langevin equation with $\gamma=\gamma_1=\gamma_2$ shows good agreements with their dependence proportional to $\delta/\gamma$ (see Fig. S2).
On the other hand, in the case where $\delta=\delta_1=-\delta_2$, there exists the opposite relationships, $\mathcal{S}\propto K_x$ and $\mathcal{K}\propto S_x$. This reference frame effectively makes a partial time-reversal operation which reverses the time evolution only in the second mechanical mode (i.e., $y_2\to-y_2$). Because this operation swaps the function of $K_x$ and $S_x$, the rotational trajectory with a real (imaginary) angle is extracted via the conditional measurement of $K_x$ ($S_x$). 
\begin{figure}
\includegraphics{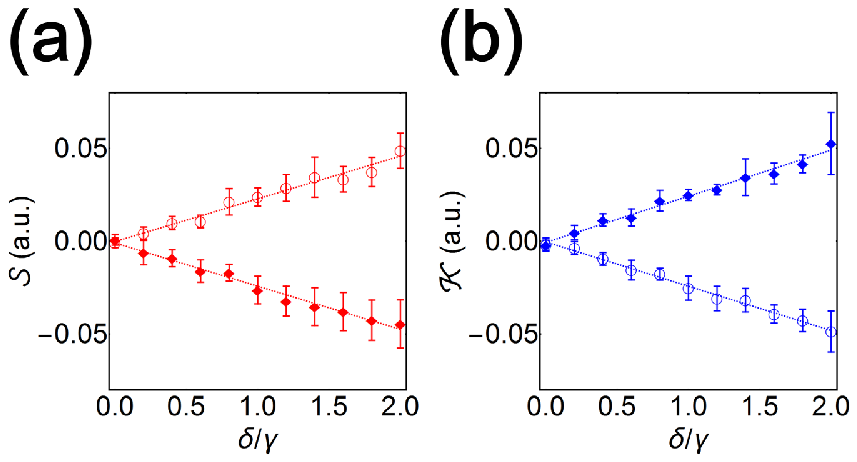}
\caption{(a) and (b) Rotational angular momenta $\mathcal{S}$ and $\mathcal{K}$ calculated with results of numerical simulation of Langevin equations with $\delta=\delta_1=\delta_2$. The square plots show rotational angular momentum with  $\eta=0.3$ in the condition $O_j/\mathrm{max}[O_j]\geq \eta$, and the circle plots show that with $\eta=-0.3$ in the condition $O_j/\mathrm{max}[O_j]\leq \eta$ where $O_j=\{S_x, K_x\}$. The error-bars represent the standard deviation for ten trials. All numerical results fitted to linear functions (shown by the dashed lines) show good agreement with the analytical expressions (\ref{Srot}) and (\ref{Krot}).}
\end{figure}

\section*{Conditional measurement with the other components of angular momenta}
Here, we show the results of the conditional measurement of $S_y$ and $K_x$, which corresponds to the real and imaginary rotation in the phase space spanned by $x_1$ and $y_2$ [Fig. S1(a) and (b)]. Moreover, the correlation coefficient matrix in the conditional measurement of SU(1,1) Schwinger angular momenta also contains the non-diagonal elements as the case of angular momenta in the SU(2) group do [Fig. S1(c)].
\begin{figure}
\includegraphics{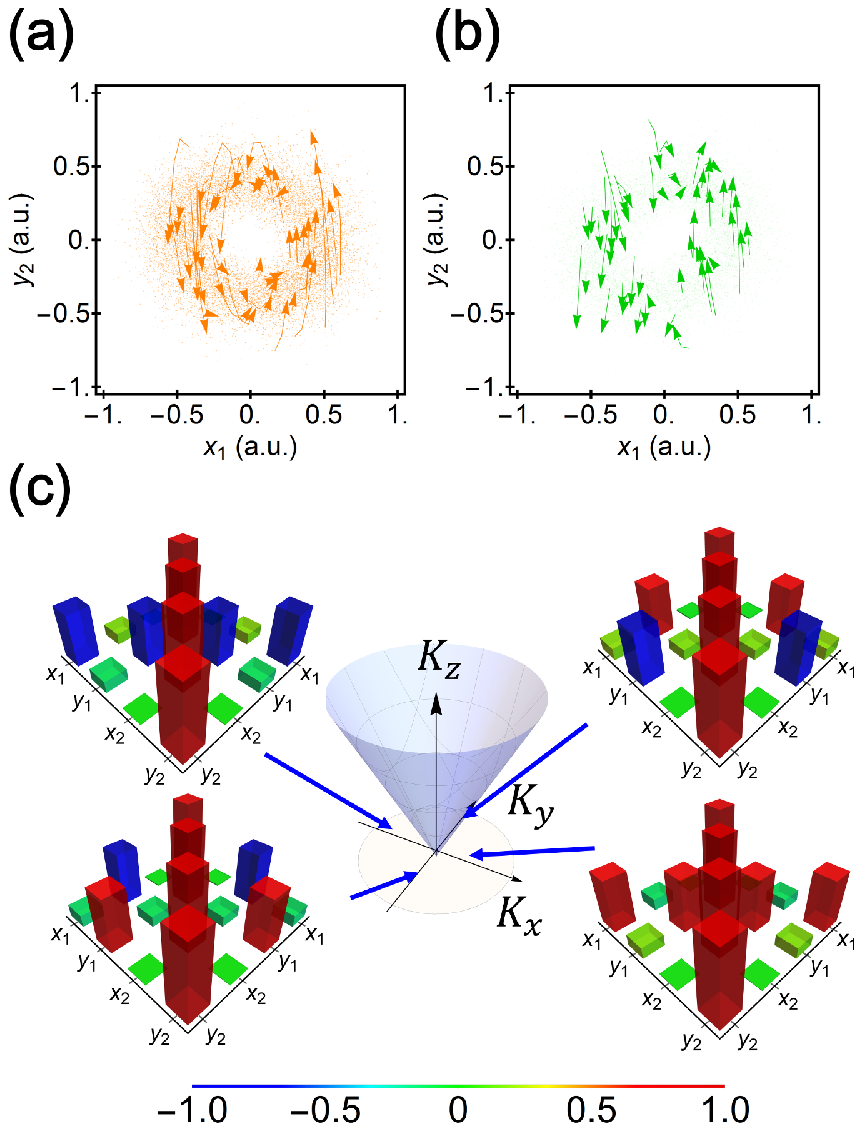}
\caption{(a) and (b) Extracted trajectory (vectors) and distributions (points) with $\eta=0.3$ for $S_y$ and $K_y$, respectively. The discontinuous trajectories are described while the Schwinger angular momenta satisfy the condition. (c) Correlation coefficient matrix with conditional measurement when $O_j=\{K_x, K_y\}$ and $\left|\eta\right|=0.2$ in SU(1,1) Schwinger angular momentum. The value including the sign of the correlation coefficient is described by the color maps.}
\end{figure}

\section*{Conditional measurement with the other double conditions}
By fixing the first condition $S_x /\mathrm{max⁡}[S_x]\leq 0.2$, the squeezed conditional distribution with respect to the angular momenta for the second conditions, $S_y$, $K_x$, and $K_y$ are compared with $\epsilon=0.1$ [Fig. S2(a)-(c)]. Obviously, the multi-modal non-Gaussian distribution appears by conditioning the angular momenta in the SU(1,1) group while the first condition is selected from the angular momentum in the SU(2) group ($S_x$). The squeezed levels along the $x_1-y_2$ axis are $-4.5\pm 0.4$, $-3.0\pm 0.4$, and $-3.0\pm 0.3$ dB in the second condition with $S_y$, $K_x$, and $K_y$, respectively. Note that the rotational trajectory in the phase space spanned by $x_1$ and $x_2$ also shows the multi-modal distribution in the case of $K_x$ and $K_y$ [Fig. S2(d)-(f)].
\begin{figure}
\includegraphics{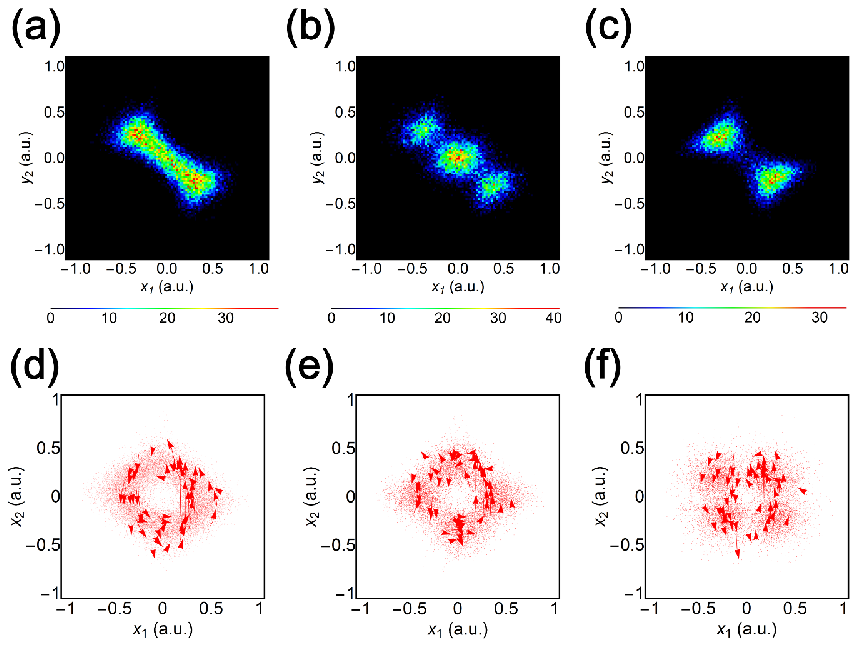}
\caption{(a), (b), and (c) Density histograms of the extracted two-mode thermal squeezed state with the double conditioning in $S_y$, $K_x$, and $K_y$. (d), (e), and (f) Conditioned distributions and trajectories extracted by the second conditions in $S_y$, $K_x$ and $K_y$, respectively. The trajectories (vector) are described while the Schwinger angular momenta satisfy each condition. }
\end{figure}
\clearpage

\bibliography{main.bib}

\bibliographystyle{unsrtnat}
\begin{thebibliography}{35}
\bibitem{schwinger}  J. Schwinger, in: L. C. Biedenharn, H. van Dam (Eds.),  Academic, New York, pp. 229?279, (1965).
\bibitem{white}A. G. White, D. F. V. James, P. H. Eberhard, and P. G. Kwiat, Phys. Rev. Lett. 83, 3103 (1999).
\bibitem{rehacek}  J. {\v{R}}eh{\'a}{\v{c}}ek, B.-G. Englert, and D. Kaszlikowski, Phys. Rev. A 70, 052321 (2004).
\bibitem{yurke}  B. Yurke, S. L. McCall, and J. R. Klauder, Phys. Rev. A 33, 4033  (1986).
\bibitem{nha}  H. Nha, and J. Kim, Phys. Rev. A 74, 012317 (2006).
\bibitem{stokes} G. G. Stokes. Trans. Cambridge Phil. Soc. 9, 399?416 (1852).
\bibitem{milione} G. Milione, H. I. Sztul, D. A. Nolan, and R. R. Alfano, Phys. Rev. Lett. 107, 053601 (2011).
\bibitem{bowen} W. P. Bowen, R. Schnabel, H-A. Bachor, and P. K. Lam, Phys. Rev. Lett. 88, 093601 (2002).
\bibitem{lassen} M. Lassen, G. Leuchs, and U. L. Andersen, Phys. Rev. Lett. 102, 163602 (2009).
\bibitem{li} D. Li, B. T. Gard, Y. Gao, C-H. Yuan, W. Zhang, H. Lee, and J. P. Dowling, Phys. Rev. A 94, 063840 (2016).
\bibitem{sankey}J. C. Sankey, C. Yang, B. M. Zwickl, A. M. Jayich, and J. G. E. Harris, Nat. Phys. 6, 707 (2010).
\bibitem{brawley} G. A. Brawley, M. R. Vanner, P. E. Larsen, S. Schmid, A. Boisen, and W. P. Bowen, Nat. Commun. 7, 10988 (2016).
\bibitem{leijssen}R. Leijssen, G. R. La Gala, L. Freisem, J. T. Muhonen, and E. Verhagen, Nat. Commun. 8, 16024 (2017).
\bibitem{okamoto}H. Okamoto, A. Gourgout, C-Y Chang,	K. Onomitsu, I. Mahboob, E. Y. Chang, and H. Yamaguchi , Nat. Phys. 9, 480 (2013).
\bibitem{faust}T. Faust, J. Rieger, M. J. Seitner, J. P. Kotthous, and E. M. Weig, Nat. Phys. 9, 485 (2013).
\bibitem{mahboob}I. Mahboob, H. Okamoto, K. Onomitsu, and H. Yamaguchi , Phys. Rev. Lett. 113, 167203 (2014).
\bibitem{pontin}A. Pontin, M. Bonaldi, A. Borrielli, L. Marconi, F. Marino, G. Pandraud, G.?A. Prodi, P.?M. Sarro, E. Serra, and F. Marin, Phys. Rev. Lett. 116, 103601 (2016).
\bibitem{karabalin}R. B. Karabalin, M. H. Matheny, X. L. Feng, E. Defa\"{y}, G. L. Rhun, C. Marcoux, S. Hentz, P. Andreucci, and M. L. Roukes, Appl. Phys. Lett. 95, 103111 (2009).
\bibitem{note}Although $\langle \tau_w (\eta) \rangle$ (here we explicitly denote it as the function of $\eta$) defines a simple time-scale for our state preparation scheme, it is not suitable to stably prepare the squeezed states with squeezing level $\xi_\mathrm{sq}(\eta)$ because of the large deviation, $\sigma(\tau_w(\eta))\sim \langle \tau_w (\eta)\rangle$. To efficiently use the two-mode thermal squeezed states for sensing or interferometry, we have to set $\tau_w(\tilde{\eta})$ so that the trajectory enters the deeper conditional window spanned by $\tilde{\eta}>\eta$ with respect to the expected squeezing level $\xi_\mathrm{sq}(\eta)$.
\bibitem{mathur}M. Mathur, I. Raychowdhury, and R. Anishetty, J. Math. Phys. (N.Y.) 51, 093504 (2010)
\bibitem{mahboob2}I. Mahboob, M. Mounaix, K. Nishiguchi, A. Fujiwara, and H. Yamaguchi , Sci. Rep. 4, 4448 (2014).
\bibitem{hvidtfelt}W. Hvidtfelt P. Nielsen, Y. Tsaturyan, C. B. Moller, E. S. Polzik, A. Schliesser, Proc. Natl. Acad. Sci. 114, 62 (2017).
\bibitem{groblacher}S. Gr\"{o}blacher, A. Trubarov, N. Prigge, G. D. Cole, M. Aspelmeyer, and J. Eisert, Nat. Commun. 6, 7606 (2015).
\bibitem{klaers}J. Klaers, S. Faelt, A. Imamoglu, and E. Togan, Phys. Rev. X 7, 031044 (2017).
\bibitem{jacobs}K. Jacobs, L. Tian, and J. Finn, Phys. Rev. Lett. 102, 057208 (2009).
\bibitem{vanner}M. R. Vanner, Phys. Rev. X 1, 021011 (2011).
\bibitem{vourdas} A. Vourdas, Phys. Rev. A 46, 442 (1992).
\bibitem{bunch}J. S. Bunch, A. M. van der Zande, S. S. Verbridge, I. W. Frank, D. M. Tanenbaum, J. M. Parpia, H. G. Craighead, P. L. McEuen, Science 315, 490 (2007).
\bibitem{will} M. Will, M. Hamer, M. Moller, A. Noury, P. Weber, A. Bachtold, R. V. Gorbachev, C. Stampfer, and J. Gttinger, Nano Lett. 17, 5950 (2017).
\bibitem{he2}R. He, X. L. Feng, M. L. Roukes, and P. Yang, Nano Lett. 8, 1756 (2008).
\bibitem{abhilash}T. S. Abhilash, J. P. Mathew, S. Sengupta, M. R. Gokhale, A. Bhattacharya, and M. M. Deshmukh, Nano Lett. 12, 6432 (2012). 
\bibitem{tsaturyan} Y. Tsaturyan, A. Barg, E. S. Polzik, and A. Schliesser, Nat. Nanotech. 12, 776 (2017).
\bibitem{ghadimi}A. H. Ghadimi, S. A. Fedorov, N. J. Engelsen, M. J. Bereyhi, R. Schilling, D. J. Wilson, and T. J. Kippenberg, Science 360, 764 (2018).
\end{thebibliography}

\end{document}